\documentclass[12pt]{article}

\newcommand{\be}{\begin{equation}}
\newcommand{\ee}{\end{equation}}

\def\P{{\bf P}}

\begin{document}


\begin{titlepage}

\begin{flushright}
{\tt LTH-1047}\\
\end{flushright} 

\vskip 2cm
\begin{center}
{\large \bf  Geometric Constructions of Two Dimensional (0,2) SUSY Theories}
\vskip 1.2cm
Radu Tatar 

\vskip 0.4cm

{\it Division of Theoretical Physics, Department of Mathematical Sciences

The University of Liverpool,
Liverpool,~L69 3BX, England, U.K.

rtatar@liverpool.ac.uk}

\vskip 1.5cm

\abstract{We consider the field theories on multiple stacks of D5 branes wrapped on four cycles of resolved/deformed conifold geometries fibered over a two torus. The central charges of the D5 branes are slightly misaligned when the branes are wrapped on various rigid holomorphic two cycles or when they have different charges with respect to a magnetic flux turned on the two torus. The  wrapped D5 branes preserve (0,2) supersymmetry in two dimensions if the Kahler moduli and the magnetic flux  are related. Our geometries are T-dual to the brane configurations considered by Kutasov-Lin and we provide a geometric interpretation for their equality between the field theory D-terms and the magnetic fluxes. We  also consider the geometric transitions for rigid holomorphic two cycles fibered over a two torus with magnetic flux and discuss the partial breaking of supersymmetry after the geometric transition.} 

\end{center}
\end{titlepage}


\section{Introduction}
The supersymmetric field theories enjoy some elegant descriptions in string theory compactifications. One successful direction 
of research studies geometric transitions which map wrapped brane setups into flux configurations, as proposed in \cite{vafa} and extended in \cite{vafa1}. The transition can also be understood by studying matrix models which allow perturbative insights into nonperturbative physics \cite{dva}. A configuration with wrapped antibranes can also provide supersymmetric configurations before and after the transition \cite{aga2}. A natural generalization to hybrid system of branes and antibranes was considered in 
\cite{aga1} to tackle the problem of D-term supersymmetry breaking.  

Soon after the geometric transition was described by studying D5 branes  wrapped on 2-cycles, a T-dual picture was proposed where the wrapped D-branes are mapped into D branes suspended between various types of NS branes \cite{rt}. The brane picture allows a lift to M-theory and the use of the MQCD approach to obtain details about the geometric transitions. The configuration of D4 and NS branes is lifted as a unique M5 brane which splits into a collection of simpler M5 branes after the geometric transition \cite{rt}. 

Recently there has been an increasing interest in using branes and geometry to study two dimensional field theories. A class of interesting theories are the chiral (0,2) SUSY theories in two dimensions. The first brane construction was proposed some time ago and involved three sets of orthogonal NS branes \cite{gu}. More recently, two dimensional (0,2) theories  emerged from compactifications of six dimensional theories on 4-manifolds with a partial topological twist \cite{guko1}. This led to the realization of some interesting two dimensional triality as an IR equivalence between three different theories \cite{guko2}. Other developments include twisted compactifications of the four-dimensional Leigh-Strassler fixed point on closed hyperbolic Riemann surfaces \cite{guko3} and a Pfaffian description \cite{guko4}.    

An alternative approach was proposed in \cite{kuta1,kuta2} utilizing brane configurations with colour D4 branes and flavour D6 branes suspended between orthogonal NS branes. The corresponding four dimensional ${\cal N} = 1$ supersymmetric field theories were further compactified on a two torus to yield (2,2) SUSY two dimensional theories. A D-term for the field theory on the D4 branes (either colour or flavour groups) and a magnetic flux on the two torus were added as extra ingredients representing rotations and displacements of various D4 branes and NS branes. This leads generically to SUSY breaking but a fine tuning for the D-term and the magnetic flux can conspire to partially preserve some supersymmetry, in particular (0,2) SUSY in two dimensions. 

In this work, our goal is to study the geometric picture arising from T-dualising the brane configuration of \cite{kuta1}. The T-duality leads to multiple stacks of  D5 branes wrapped on $\P^1$ cycles or non-compact holomorphic cycles. To obtain a two dimensional theory, we fibre the resolved conifold geometries over a two torus and reinterpret the setup as wrapped D5 branes on $\P^1$ fibres over the $T^2$. After turning on a D-term on the $\P^1$ fibre 
(making $\P^1$ cycle rigid), the central charges of the branes become misaligned \cite{aga1} which potentially leads to supersymmetry breaking. On the other hand, turning on a magnetic flux through the two torus could also break supersymmetry. We discuss how these two types of SUSY breaking can compensate each other and partially preserve the SUSY for branes wrapped on 4 cycles inside $SU(4)$ structure manifolds (when an extra NS flux is present). We consider  the SUSY condition for wrapped D5 branes on 2-cycles and 4-cycles of $SU(3)$ and $SU(4)$ holonomy manifolds derived in \cite{moore} and replace the Kahler 2-form $J$ with its complexified version. The SUSY condition becomes an equality between the Kahler form and the magnetic flux through the two torus base, which represents a geometric interpretation of the equality between the D term and the magnetic flux proposed in  \cite{kuta1,kuta2}. 
  
In section 2, we start by reviewing the geometric D-term SUSY breaking considered in 
\cite{vafa1,aga1}. For a single stack of D5 branes, a SUSY configuration can be obtained even in the presence of D-terms/rigid cycles but this is not true for D5 branes wrapped on arbitrary rigid $\P^1$ cycles or noncompact 2-cycles. We also consider the boost supergravity solution described in \cite{mm} and discuss the gauge coupling constant on wrapped D5 branes. In section 3, we review the proposal  of  \cite{int1} to build Calabi-Yau fourfolds as resolved/deformed conifolds fibered over a genus $g$ base and we restrict to the case $g=1$. 

In section 4 we consider the unbroken supersymmetry condition for D5 branes wrapped on rigid $\P^1$ cycles fibered over $T^2$ with magnetic flux. Our main claim is that the condition of SUSY preservation is satisfied when D5 branes wrap Kahler calibrations and the Kahler moduli and the magnetic field are related, reproducing  the condition derived in \cite{kuta1}. In section 5 we consider the geometric transition inside the SU(4) structure manifolds, in the presence of  nonzero D-terms and magnetic fluxes. After the transition, the colour D5 branes are replaced by fluxes through various $S^3 \times S^1$ cycles and the gluino condensates are equal to the integrals of the holomorphic 4-form on such cycles. The flavour degrees of freedom lie on D5 branes wrapped on noncompact 2-cycles, which remain unchanged during the geometric transition. The cancellation between the global symmetry D-term  and the magnetic field remains valid during the transition and assures SUSY preservation. 
 
\section{D-terms for Wrapped D5 Branes}

\subsection{The Geometry of D-terms}

We start by reviewing the geometric interpretation of the  D-terms for
${\cal N} = 1, d=4$ field theories. Consider a resolved conifold and wrap some D5 branes on the non-rigid $\P^1$ cycle. The gauge coupling is 
\begin{equation}
\frac{4 \pi}{g^2_{YM}}= \frac{b_{NS}}{g_s}
\end{equation}
where $b_{NS}$ is the integral over $\P^1$ of the NS two-form field on the D5 branes. 

In addition, we can turn a small nonzero Fayet-Iliopoulos parameter $\xi$ for the $U(1)$ center of the gauge group which contributes to the Lagrangian with a term 
$\sqrt{2} \xi  \mbox{Tr} \mbox{D}$. Its  geometric interpretation was provided in \cite{aga1}, where it was associated to turning on the real part $j$ of the complexified Kahler class of the $\P^1$ cycle. The central charge for wrapped D5 branes is the integral of the complexified Kahler form
\begin{equation}
Z = \int_{S^2} (J + i B_{NS}) = j + i b_{NS},
\end{equation}
where $j$ is related to $\xi$ by
\begin{equation}
\xi = \frac{j}{4 \pi g_s}.
\end{equation}
For $j \ne 0$, the phase of the central charge is modified and the supersymmetry appears to be broken due to the presence of the Fayet-Iliopoulos term. Nevertheless,  this is not necessarily true for any $j \ne 0$ \cite{vafa1}. For a single set of wrapped branes on a $\P^1$, the theory has an alternative SUSY description with a  bare coupling constant related to the quantum volume of the resolution $\P^1$ cycle as
\begin{equation}
\label{gaugecoupling}
\frac{4 \pi}{g^2_{YM}}= \frac{\sqrt{b_{NS}^2+ j^2}}{g_s}.
\end{equation}
For a product group obtained on several stacks of D5 branes wrapped on different $\P^1$ cycles,  we have the freedom to turn different values for $j$ on  each of the $\P^1$ cycles. For two stacks of branes wrapped on $\P^1$ cycles with $j_1 \ne j_2$, the central charges have different phases, they cannot align and the supersymmetry is broken \cite{aga1}.

\subsection{The Supergravity Interpolating Solution}

The variation of the parameters $J$ and $B_{NS}$ for the wrapped D5 branes was studied in  supergravity by many authors \cite{minagra,ds1,mm,kem}. \cite{minagra,ds1} considered a flow between a Maldacena-Nunez solution \cite{mn} and a Klebanov-Strassler solution \cite{ks}. The Maldacena-Nunez solution corresponds to large values for $J$ and
zero $B_{NS}$ whereas the Klebanov-Strassler solution is valid for zero $J$ (fractional branes) and non-zero 
$B_{NS}$.  The solution involves a reduction of 10 dimensional spinors $\epsilon_i,i=1,2$ to six dimensional spinors $\eta_{+}^i, i=1,2$ which are related to the SU(3) invariant spinors $\eta_{+}$ as \cite{minagra}
\be
\label{sugra}
\eta_{+}^1 = \frac{1}{2} (\alpha + \chi) \eta_+;~~ \eta_{+}^2 = \frac{1}{2i} (\alpha - \chi) \eta_+  
\ee
The choice $\alpha = 0$ (or $\chi = 0$) corresponds to the Maldacena-Nunez solutions and 
$\alpha = \pm i \chi$ corresponds to the Klebanov-Strassler solution. The interpolating solution is parametrised by a phase $\omega$ related to $\chi$ and $\alpha$ as $\chi = i~\mbox{sin} (\omega/2);~ \alpha = \mbox{cos} (\omega/2)$. The relation (\ref{sugra}) becomes 
\be
\label{sugra1}
\eta_{+}^1 = i e^{i \omega} \eta_{+}^2.
\ee
We now compare (\ref{sugra1}) with the supersymmetry condition obtain for D5 branes wrapped on a 2-cycle of an $SU(3)$ structure manifold (when NS flux is present). The corresponding relation between $\eta_{+}^i$ was considered in \cite{moore} for $SU(3)$ holonomy and extended in \cite{uc} to $SU(3)$ structure manifolds as
\be
\label{moore1}
\eta_{+}^1 = - e^{- i \rho}  \eta_{+}^2.
\ee 
where $\rho$ is a geometric parameter. We see that the supergravity parameter $\omega$ and $\rho$ are related as $\omega = \pi/2 - \rho$. 

The flow of \cite{minagra} was reinterpreted in \cite{mm} as starting with D5 branes with no NS flux and performing a boost which provides some NS flux, after a series of S and T dualities.  This approach was subsequently used by \cite{cdt,kem} to describe wrapped D5 branes on a resolved conifold. It was argued that the SUSY preservation implies that the D5 branes should wrap a cycle inside a non-Kahler deformation of the resolved conifold. When the dilaton is constant, the IIB configuration of \cite{mm,kem} implies the following form for the RR and NS 3-forms:
\be
\label{fluxes}
H_{RR} = \mbox{cosh}~\beta *_{6} d J,~~~H_{NS} = - \mbox{sinh}~\beta~d J,
\ee
where the Hodge star is with respect to the non-Kahler metric on the resolved conifold. The supersymmetry is preserved if the $G_3 = H_{RR} - i e^{\phi} H_{NS}$ flux is of (2,1) form. For a complex internal manifold,  the dilaton is constant $\phi = \phi_0$ and the complex structure is provided by 
\be
\label{gamma}
\gamma =  e^{\phi_0} \mbox{cotanh}~\beta  
\ee
where $\gamma$ was introduced in \cite{kem} in the definition of the complex forms needed to separate the (2,1) and (1,2) pieces of the fluxes.  
We repeat the steps of \cite{mm} in case of two stacks of D5 branes.  We start with two sets of D5 branes wrapped on two $\P^1$ cycles  and compactify three extra coordinates of the D5 branes into 
a three torus and T-dualize along them to obtain two sets of D2 branes wrapped on $\P^1$ cycles. We lift this configuration to M theory and get two stacks of M2 branes. The configuration is compactified on a 7 dimensional manifold whose base is the resolved geometry with two $\P^1$ cycles. We now perform the boost of \cite{mm}:
\be
\label{boosting}
t  \rightarrow \mbox{cosh} \beta t -  \mbox{sin} \beta  x_{11},~  
x_{11}  \rightarrow - \mbox{sinh} \beta t -  \mbox{cos} \beta  x_{11}.
\ee
After reducing back to type IIA and reversing the three T-dualities, we reach a type IIB solution with two stacks of D5 branes wrapped on $\P^1$ cycles. The calibration condition becomes
\be
\label{2form}
B_{NS} = \mbox{sinh} \beta~~e^{- 2 \phi} J.
\ee
When integrating (\ref{2form}) over the two $\P^1$ cycles for a constant dilaton,  
the supersymmetry condition implies that $b_i = \int_{\P^1_i} B_{NS}$ and  $j_i = \int_{\P^1_i} J$ are related as 
\be
\label{bjrelation}
b_i = \mbox{sinh} \beta~~e^{- 2 \phi_0} j_i \rightarrow \frac{b_1}{j_1} = \frac{b_2}{j_2}
\ee
On the other hand, the central charges on the two stacks of D5 branes are $j_i + i b_i$ so the equality 
(\ref{bjrelation}) implies that the phases of the central charges are equal. For generic values of $j_i, b_i$, the condition (\ref{bjrelation}) is not satisfied and the supersymmetry is broken.  

\subsection{Gauge Coupling Constant on the Wrapped D5 branes}
The interpolating supergravity solution (\ref{sugra}), (\ref{sugra1}) constructed in \cite{minagra} involves a parameter $\omega$ which is related to the boosting parameter $\beta$ as
\be
\mbox{cos}~\omega = - \mbox{tanh}~\beta~e^{\phi}.
\ee
We now consider the gauge coupling constant on D5 branes wrapped on $\P^1$ cycles. In \cite{mm}  this was related to $\omega$  by looking at the superpotential after the geometric transition. Here we discuss the case before the geometric transition.  If we replace $J$ by its complexified version $J + i B$, reconsider the compactification manifold as an $SU(3)$ structure manifold  and use (\ref{moore1}), we get the second condition to preserve SUSY for a D5 brane wrapped on the $\P^1$ cycle \cite{moore}
\be
\label{complexcoupling}
(J + i B_{NS}) e^{- i \rho} = \mbox{vol}_{\P^1}
\ee
The angle $\rho$ is a parameter, as defined in equation (7). In order to obtain a real left hand side in equation (\ref{complexcoupling}), $J + i B_{NS}$ should have a phase equal to $\rho$ so we can identify the  parameter as $J~\mbox{tan}(\rho) = B_{NS}$.  

After integrating over $\P^1$, the left hand side of (\ref{complexcoupling}) becomes 
\be
\label{ggc}
j \mbox{cos} \rho + b \mbox{sin} \rho = \sqrt{j^2 + b^2} 
\ee
which is the inverse of the gauge coupling constant for D5 branes wrapped on a rigid $\P^1$.

\subsection{T-dual picture}

What happens when we perform a T-duality along the angular direction of the $\P^1$ cycle? The singular lines inside the resolved conifold are replaced by two orthogonal NS branes and the value of the  D-term maps into some extra  separation between the NS branes and a rotation of the D4 branes. If the NS branes extend in the directions (012345) and (012389), $j=0$ corresponds to NS branes being separated only in the $x^6$ direction. 
$j \ne 0$ adds an extra displacement of the NS branes in the direction $x^7$ and a rotation of the  D4 branes in the $(x^6, x^7)$ plane. 

When starting with two stacks of D5 branes wrapped on two rigid $\P^1$ cycles, the T-dual configurations contains three NS branes separated in the $(x^6,x^7)$ plane with two stacks of D4 branes between them. For different values of $j_1,j_2$, the two stacks of D4 branes are rotated by different angles, which signals SUSY breaking.  When one of the  $\P^1$ cycles is replaced with a non compact holomorphic two cycle, we get a gauge group with flavours. In the T-dual picture, flavour groups are represented by semi-infinite D4 branes, rotated in the $(x^6, x^7)$ plane when $j \ne 0$.  Turning on a Fayet-Iliopoulos term implies a  rotation of the flavour (semi-infinite) D4 branes with respect to the colour (finite) D4 branes in the $(x^6,x^7)$ plane and the supersymmetry is generically broken. The configuration with rotated semi-infinite D4 brane and unrotated finite D4 branes maps into the D6 brane picture of \cite{kuta1} via a Hanany-Witten brane creation effect \cite{hw}.

Supersymmetry is broken when the relation (\ref{bjrelation}) is not satisfied, as it happens for arbitrary $j_i,b_i$. To remove the requirement (\ref{bjrelation}), we uplift the resolved conifold geometry to a resolved conifold fibration over a two torus $T^2$. The D5 branes wrap a $\P^1$ fibration over $T^2$ and we add some magnetic flux on the $T^2$ base which combines with the rigidity parameter $j$ to provide a SUSY configuration. Such constructions (without magnetic flux and D-terms)  were introduced in 
\cite{int1} and we consider these geometries in the next section. 

\section{D5 branes wrapped on 4-cycle inside CY 4 folds}

We briefly review the set-up  of resolved/deformed conifold fibrations over $T^2$ proposed in \cite{int1}. We consider D5 branes wrapped on 4-cycles inside Calabi-Yau 4-folds described as fibrations over genus $g$ surfaces.  
We start with a conifold represented by the equation
\begin{equation}
x_1 x_2 - x_3 x_4 = 0,
\end{equation}
in terms of the complex coordinates $x_i, i=1,\cdots,4$. When the coordinates $x_i$ are line bundles ${\cal L}_i$ over a curve $C$, the eight dimensional manifold becomes a Calabi-Yau fourfold $X_s$ in the five dimensional  complex variety ${\cal L}_1 \oplus {\cal L}_2 \oplus {\cal L}_3 \oplus {\cal L}_4 \rightarrow C$.

The singular Calabi-Yau fourfold $X_s$ can be made smooth by either a small resolution (a fourfold denoted $X_r$) or by a deformation along the curve $C$ (a fourfold denoted $X_d$). 

\subsection{Resolved Conifold Side}

Consider  $X_r$ an  $O(-1) \oplus O(-1) \rightarrow \P^1$ fibration over a genus g curve. 
The $\P^1$ fibration over $C$ gives rise to a compact two complex dimensional surface $S$ with Euler characteristic $\chi(S) = 4 - 4 g$. If the line bundle 
${\cal L}_1 \otimes {\cal L}_4^{-1}$ has degree $n$, the volume of $S$ is
\be
\label{volume}
Vol(S) = \frac{n}{2} (J^{F})^2 + J^{F} J^{C}
\ee
where $J^{F}$ and $J^{C}$ measure the volumes of the $\P^1$ fiber and of the curve $C$. If we wrap D5 branes on $S$, we get a two dimensional field theory with reduced supersymmetry.

There are some other two cycles in the $O(-1) \oplus O(-1) \rightarrow \P^1$ fibre. The small resolution is covered by two copies of ${\bf C}^3$ with coordinates $Z,X,Y$ and $Z',X',Y'$ respectively. One can define two types of non-compact holomorphic cycles
$\tilde{B}_1:~Y=0, X=m$ or $\tilde{B}_2:~Y'= 0, X' = M$. If we wrap D5 branes on the non-compact holomorphic cycles, they will correspond to massive flavour (if wrapped on $\tilde{B}_1$) or flavour with expectation value (if wrapped on  $\tilde{B}_2$). The $\tilde{B}_1$ fibration over $C$ is a non-compact two complex dimensional surface $B_1$ and the $\tilde{B}_2$ fibration over $C$ is a non-compact two complex dimensional surface $B_2$. If we wrap D5 branes on $B_1$ or $B_2$, we obtain two dimensional flavour fields. 

\subsection{Deformed Conifold Side}
  
After a geometric transition on the fibre, a Calabi-Yau four fold $X_d$ is obtained by deforming the singular fibre of $X_s$ into a deformed conifold with a deformation parameter $\epsilon$:
\be
\label{dcon}
 x y - u v =\epsilon.
\ee
The singular point is replaced by a 3-cycle $S^3$ and its Poincare dual $P$. There are several types of non-trivial four-cycles inside $X_d$:

$\bullet$  $2 g$ four-cycles $D_n, n= 1,\cdots, 2 g$ of topology $S^1 \times S^3$ generated by transporting the $S^3$ fibers  along the non-trivial $2 g$ cycles of the base. 

$\bullet$  $2 g - 3$ four cycles of topology $S^4$. 

\subsection{Our case: Two Torus (g=1)}

In this work we consider a two torus base and choose the value of $n$ is (\ref{volume}) to be zero. This simplifies the problem because:

$\bullet$ the volume of the surface $S$ in the resolved conifold is $J^{F} J^{C}$.

$\bullet$ there is no cycle of topology $S^4$. For each deformation $S^3$ cycle, we get two $S^1 \times S^3$ cycles which we denote by $D_1, D_2$ and their  Poincare duals $\tilde{D}_1, \tilde{D}_2$.

The 4-cycles $B_1$ and $B_2$ which correspond to massive flavours or flavours with expectation values also exist in the deformed geometry.

\section{Supersymmetric Configurations with D-terms and Magnetic Fluxes through $T^2$}

Consider D5 branes wrapped on various $\P^1$ or $\tilde{B}_i$ fibrations over a two torus. We are interested to have a 
$U(N_c) \times U(N_f)$ theory with $U(N_c)$ a gauge group and $U(N_f)$ a global flavour group. In this work we only consider the case of a D-term for a $U(1)$ subgroup of $U(N_f)$. In the absence of magnetic  flux through the two torus, the resulting 8-dimensional manifold inherits the $SU(3)$ holonomy from the resolved conifold. The theory on each D5 branes is  2 dimensional with a (2,2) supersymmetry. The four cycle  $T^2 \times \P^1$ is holomorphic and the coupling constant of the two dimensional field theory obtained on each D5 branes wrapped on $T^2 \times \P^1$ is 
\be
\label{nof}
\frac{1}{g_{(2,2)}} = \frac{b_{NS} A_{T^2}}{g_s}
\ee
where $b_{NS}$ is the integral of $B_{NS}$ through the non-rigid $\P^1$ cycle and 
$A_{T^2}$ is the area of the two torus. 
Due to the presence of the NS flux, the 4-cycle $\P^1 \times T^2$ is a generalized holomorphic cycle embedded in a 
$SU(3)$ structure manifold. 

In the absence of magnetic fluxes and in the limit 
$\omega = 0$ for the four dimensional $N=1$ SUSY theory, the relation (\ref{sugra1}) becomes
\be
\eta_{+}^1 = i \eta_{+}^2
\ee
which is just the limit $\theta = 0$ of the unbroken supersymmetry condition for D branes wrapped on a four-cycle \cite{moore}. We conclude that the $\theta = \omega = 0$ solution (\ref{nof}) fits the SUSY condition for 
D5 branes wrapped on 4-cycles inside $SU(3)$ structure manifolds. We now vary $\theta$ and $\omega$. When $\theta$ is arbitrary for $\omega = 0$, the supersymmetry is preserved and remains (2,2) in 2 dimensions. For arbitrary $\theta$ and $\omega$, the supersymmetry is generically broken. 

\subsection{SUSY and Magnetic Fluxes}

We first consider the case of a constant $\omega = 0$ and an arbitrary $\theta$. This correspond to having $J=0$ (infinite boost) and a vector potential $A_2 = M x^1$ on the two torus. If the D5 branes are only wrapped on a two torus with magnetic flux, the SUSY conditions imply a relation between the spinors $\eta^{i}_{+}$ containing a parameter $\theta$:~~$\eta_{+}^1 = - e^{- i \theta} \eta_{+}^2$ and one involving 2 forms
\be
\label{susy0}
\mbox{vol}_{T^2} + i M = e^{i \theta} \frac{\sqrt{|g+M|}}{\sqrt{|g|}} \mbox{vol}_2
\ee
where $\mbox{vol}_{T^2}$ is the volume form of the two torus and $M$ is a two form.

We now consider wrapping the D5 branes on a trivial $\P^1$ fibre over $T^2$, with $B_{NS}$ on $\P^1$. The 
trivial $\P^1$ fibre over $T^2$ is a 4-cycle. For $\omega = 0$ (infinite boost), we have $J = 0, B_{NS} \ne 0$. The right hand side of (\ref{susy0}) becomes the 
volume of a four cycle whereas the left hand side of (\ref{susy0}) 
is multiplied by $i B_{NS}$. We insert the factor $i$ in the relation between the $\eta_{+}^i$, which becomes 
$\eta_{+}^1 = i e^{- i \theta} \eta_{+}^2$. This is exactly the unbroken supersymmetry condition for D branes wrapped on 4-cycles of SU(3) holonomy manifolds \cite{moore} (modified to SU(3) structure 
in the presence of $B_{NS})$. The supersymmetry condition (\ref{susy0}) becomes
\be
\label{susy1}
B_{NS} \wedge (\mbox{vol}_{T^2} + i M) = e^{i \theta} \frac{\sqrt{|g+M|}}{\sqrt{|g|}} \mbox{vol}_4
\ee
where $\mbox{vol}_4$ is the volume form of the $\P^1$ fibration over $T^2$. 

We introduce a phase $\sigma$ as a function of the magnetic flux $M$:
\be
\label{fluxt}
M  = \mbox{tan}~(\sigma)~\mbox{vol}_{T^2}.
\ee

We integrate (\ref{susy1}) over the $\P^1$ fibration over the $T^2$ and denote
\be
b_{NS} = \int_{P^1} B_{NS},~~I_4 = \int \frac{\sqrt{|g+M|}}{\sqrt{|g|}} \mbox{vol}_4,~
A_{T^2} = \int_{T^2} \mbox{vol}_{T^2},
\ee
As $M$ is a constant flux on the two torus, $\int_{T^2} M = M~A_{T^2}$, where $M$ is now a number. 
The result of integrating (\ref{susy1}) is then
\be
\label{flut1}
b_{NS} A_{T^2} \sqrt{1+ M^2} e^{i \sigma} = e^{i \theta} I_4.
\ee
This relation is satisfied if $\sigma = \theta$ and
\be
\label{susy11}
b_{NS} A_{T^2} \sqrt{1 + M^2} = I_4.
\ee 
This enables us to identify the SUSY parameter $\theta$ with $\sigma$ of (\ref{fluxt}) 
and the gauge coupling constant with the inverse of $b_{NS} A_{T^2} \sqrt{1 + M^2}$. 
An argument for this value of the coupling constant was provided in \cite{lust} where the coupling constant for the gauge theory on a D-brane wrapped on a torus with area $A_{T^2}$ and in the presence of a magnetic flux $M$ was shown to be  $A_{T^2}\sqrt{1 + M^2}$.  This is exactly the interpretation of (\ref{susy11}) after a further compactification on a non-rigid $\P^1$.  (\ref{susy11}) therefore provides  the (2,2) two dimensional coupling constant on D5 branes wrapped on the direct product $T^2 \times \P^1$, with magnetic flux $M$:
\be
\label{nof1}
\frac{1}{g_{(2,2)}} = \frac{b_{NS} A_{T^2}  \sqrt{1 + M^2} }{g_s}
\ee
The change in the coupling constant determined by the magnetic flux can also be understood in a T-dual picture with D branes, after the T-duality  is taken along one of the directions of $T^2$. The D4 branes are replaced by D3 branes with a tilting in the $(1,2)$ plane due to the magnetic field. The 2-dimensional coupling constant is inverse proportional to the length of D3 brane i.e. proportional to  $1/\sqrt{1 + M^2}$.
 
\subsection{SUSY configurations with Magnetic Flux and Rigid Cycles}

We consider the general solution for arbitrary $\theta$ and an arbitrary boosting parameter 
$\omega$. This corresponds to allowing some arbitrary $J$ and $M$. When starting with a group 
$U(N_c) \times U(N_f)$, the magnetic flux is chosen such that the gauge group is broken to 
$U(N_c) \times SU(N_f/2)^2 \times U(1)$.
We want zero entries for the $N_c \times N_c$ block as the $U(N_c)$ fields are not charged under the magnetic flux. 

The resolved conifold is now non-trivially fibered over $T^2$ and we deal with a Calabi-Yau 4-fold. We have various types of stable 4-cycles inside Calabi-Yau 4-folds. One type of stable 4-cycles are the Kahler calibrations which are calibrated by $J^2$ and are complex submanifolds. The second type of 4-cycles are the Lagrangian submanifolds $L$ which are calibrated by $\alpha_{\psi} = \mbox{Re} (e^{i \psi} \Omega)$ where $\Omega$ is the holomorphic (4,0) form and $\psi$ is a phase. The most general calibration is the Cayley calibration when the 4-cycles are calibrated by $J^2 + \mbox{Re} (e^{i \psi} \Omega)$. The Cayley calibrations for wrapped D-branes were first used in \cite{becker,becker1}. 

In this work we consider $S$ is a non-trivial $\P^1$ or $\tilde{B}$ fibration over $T^2$. It is  a compact complex submanifold, as introduced in \cite{int1}. This implies that we deal with Kahler calibrations on which  $\mbox{Re}(f^{*}(e^{i \psi} \bar{\Omega}))$ is zero. The conditions for unbroken SUSY on D5 branes wrapped
on Kahler calibrations are \cite{moore}
\be
\label{relsu41}
\eta_{+}^1 = i e^{- i(\theta + \phi)} \eta_{+}^2.
\ee
and 
\be
\label{relsu42}
(J + i B_{NS}) (\mbox{vol}_{T^2} + i M) = e^{i (\theta+\phi)} 
\frac{\sqrt{|g+M|}}{\sqrt{|g|}} \mbox{vol}_4.
\ee
The conditions (\ref{relsu41}), (\ref{relsu42}) for D5 branes wrapped on a 4-cycle of an $SU(4)$ structure manifold have two angular parameters $\theta$ and $\phi$ which we want to relate to the parameters 
\be
B_{NS} = \mbox{tan}(\rho)~J,~M = \mbox{tan}(\sigma)~\mbox{vol}_{T^2}.
\ee  
To do this, we start  with D5 branes wrapped on a 2 torus
with magnetic flux which requires $\eta_{+}^1 = - e^{- i \theta} \eta_{+}^2$ and fix the $\theta$ parameter to be equal to $\sigma$ such that  $M = \mbox{tan} (\theta)~\mbox{vol}_{T^2}$. We then wrap the D5 branes on an extra $\P^1$ cycle such that the $\P^1$ fibre over the two torus is a Kahler calibration inside an $SU(4)$ holonomy manifold, as in \cite{int1}.   

In case of extra wrapping on a rigid 2-cycle, we multiply $\mbox{vol}_{T^2} + i M$ by $J + i B_{NS}$ and the relation between $\eta_{+}^i, i=1,2$  changes from $\eta_{+}^1 = - e^{- i \theta} \eta_{+}^2$ (D5 branes on 2 cycle) to $e^{i \rho} \eta_{+}^1 = - e^{- i \theta}  \eta_{+}^2$ (D5 branes on 4-cycle). The factor $e^{i \rho}$ is the phase of $J + i B_{NS}$ and is extracted once we consider that the quantum volume of the $P^1$ cycle is
$\sqrt{j^2+b^2}$. The relation between $\eta_{+}^i$ can be rewritten as 
\be
\label{relsu43}
\eta_{+}^1 = i e^{- i (\theta + \rho) + i \pi/2} \eta_{+}^2.
\ee
When comparing (\ref{relsu41}) and (\ref{relsu43}) we see that, besides $\theta = \sigma$, we can also identify 
$\phi = \rho - \pi/2$. The reality condition $\theta + \phi = 0$ becomes $\sigma = \pi/2 - \rho$, which implies
\be
\label{relkuta}  
\mbox{tan}~\rho  = \mbox{cotan}~\sigma \rightarrow 
B_{NS} \wedge M = J \wedge \mbox{vol}_{T^2}, 
\ee
which is the geometric version of the equality between the D-term and the magnetic flux considered in \cite{kuta1}.  

The supersymmetric condition also requires
\be
A_{T^2}\sqrt{j^2 + b_{NS}^2}  \sqrt{1 + M^2} = I_4
\ee
which implies that the 2-dimensional gauge coupling for the (0,2) gauge theory is 
\be
\frac{1}{g_{(0,2)}} =  A_{T^2} \frac{\sqrt{j^2 + b_{NS}^2}  \sqrt{1 + M^2}}{g_s}
\ee
In the above discussion, we have considered that the four cycle wrapped by the D5 branes is a $\P^1$ fibre over $T^2$. We can also take the limit when the cycle $\P^1$ is replaced by a non-compact holomorphic cycle. In this case we get flavour D5 branes wrapped on noncompact cycles. The SUSY compatibility between wrapped D5 branes remains the same as in
(\ref{relkuta}). 

\section{Geometric Transition with D-terms and Magnetic Fluxes}
We  now consider the geometric transition \cite{vafa} in the presence of rigid 2-cycles  and magnetic fluxes.  
The geometric transition between resolved and deformed geometries for pure gauge theories starts with $N_c$ D5 branes wrapped on a resolved conifold, continues with shrinking the  $\P^1$ cycle and replacing it with an $S^3$ cycle with a size equal to the field theory gluino condensate
\begin{equation}
\label{gluino}
S = \int_{S^3} \Omega_{3}
\end{equation}
where $\Omega_{3}$ is the holomorphic 3-form on the deformed conifold.
The colour D5 branes disappear and are replaced by $N$ units of Ramond-Ramond flux through the $S^3$ cycle
\begin{equation}
\int_{S^3} H^{RR} = N
\end{equation}
There are others quantities which map from the resolved conifold side to the deformed conifold side, involving $P$, the Poincare dual to $S^3$. The bare gauge coupling map is
\begin{equation}
\int_{\P^1} B_{NS}  \leftrightarrow \int_{P} H_{NS}
\end{equation}
and the Fayet-Iliopoulos D-term map is \cite{aga1}
\begin{equation}
\int_{\P^1} J    \leftrightarrow \int_{P} d J
\end{equation}
A nonzero value of $d J$ in the deformed geometry implies the existence of some non-zero torsion classes, a set-up studied in detail in \cite{mcg}.
On the other hand, if fundamental flavours are present, they live on D5 branes wrapped on non-compact holomorphic 2-cycles which survive the geometric transition. In this case $J \ne 0$ on the non-compact 2-cycle before the geometric transition maps into $J \ne 0$ on the non-compact 2-cycle after the geometric transition. There is no $d J \ne 0$ contribution from the surviving non-compact holomorphic 2-cycles. 

\subsection{Breaking SUSY with Fluxes and Noncompact cycles}
We saw in the resolved conifold geometry that the supersymmetry is broken if D5 branes wrap 2-cycles with different values for $j$. In particular, the SUSY is broken when the $N_c$ colour D5 branes wrap a non-rigid  $\P^1$ cycle and the $N_f$ D5 branes wrap a rigid  noncompact holomorphic 2-cycle. How do we translate this statement into the deformed conifold side?
 
Consider the deformed conifold configuration with $N_c$ units of RR flux through the $S^3$ cycle (coming from the $N_c$ D5 branes wrapping a $\P^1$ cycle) and a noncompact 2-cycle with $N_f$ D5 branes wrapped on it. This represents the strong coupling limit of the $SU(N_c)$ field theory with $N_f$ fundamental flavours. To discuss the SUSY breaking, we consider this configuration as a limit of a geometry describing the strongly coupled $SU(N_c) \times SU(N_f)$ gauge theories when the gauge coupling of $SU(N_f)$ goes to zero. If $S_1, S_2$ are the gluino condensates for $SU(N_c) \times SU(N_f)$, this limit implies that $S_2 \rightarrow 0$.  
For two $S^3$ cycles with sizes $S_1, S_2$, the prepotential is
\be
\label{prepot}
2 \pi i {F}_0 = \frac{1}{2} S_1^2 \mbox{log} (\frac{S_1}{\Lambda_0^2} - \frac{3}{2}) + 
\frac{1}{2} S_2^2 \mbox{log} (\frac{S_2}{\Lambda_0^2} - \frac{3}{2}) - S_1 S_2 \mbox{log} \frac{a}{\Lambda_0}
\ee
whose derivatives are the two B-periods $\Pi_1, \Pi_2$ of the geometry. In the limit $S_2 \rightarrow 0$, the contribution of $\Pi_1$
to the effective superpotential is the usual one for a decoupled $SU(N_c)$ gauge theory
\be
N_c(3 S_1 \mbox{log} \Lambda_0 + S (1 - \mbox{log} S_1)),
\ee
whereas the contribution of $\Pi_2$ becomes
\be
\label{ncomp}
S_2 \mbox{log} (\frac{a}{\Lambda_0}).  
\ee
We recognize the quantity (\ref{ncomp}) as the additional superpotential coming from the contribution of the D5 branes wrapped on noncompact 2-cycles
in the deformed geometry \cite{rt,vafa1}. $a$ is either the mass of the flavours \cite{vafa1} or their expectation value 
\cite{rt}. The geometry with nonzero $S_1, S_2$ continuously deforms into its $S_2 \rightarrow 0$ limit with one $S^3$ cycle and a noncompact 2-cycle. 

We can take a similar $S_2 \rightarrow 0$ limit when we start with a geometry with two  $S^3$ cycles of non vanishing sizes with $d J \ne 0$ on their Poincare duals. As considered in \cite{aga1}, in this case the critical points of the tree-level effective superpotential correspond to values for  $S_1, S_2$  containing the factors 
\be
\label{s1s2}
(\frac{a}{\Lambda})^{N_i \mbox{cos}~(\theta_{12})}, i=1,2
\ee
 where $\theta_{12}$ is the relative 
phase between the central charges $Z_i,i=1,2$ of the $SU$ groups.  The relative phase originates from the terms
\begin{equation}
\label{cccc}
\int_{P_{i}} d J/g_s = j_i/g_s;~~j_1 \ne j_2 .
\end{equation} 

In the limit $S_2 \rightarrow 0$, the cycle $S^3_2$ is replaced by a holomorphic noncompact 2-cycle and a nonzero value of $d J$ on  $P_{2}$ maps into a nonzero value of $J$ on the noncompact 2 cycle $\tilde{B}_2$.  
$\theta_{12}$ remains the relative phase between the central charges but now originates from the terms
\begin{equation}
\label{cccc1}
\int_{P_{1}} d J/g_s = j_1/g_s;~~\int_{\tilde{B}_2} J/g_s = j_2/g_s.
\end{equation} 
The 3-cycle $P_{1}$ is the Poincare dual to $S^3_1$ and should not be confused with the two cycle $\P^1$. 
If $j_1 \ne j_2$ in (\ref{cccc1}), the supersymmetry is broken. The particular case we are interested is when
$j_1 = 0$ and $j_2 \ne 0$ which occurs when a D-term is turned only for the flavour group. We see that SUSY is 
broken in this particular case.  

\subsection{Reduction on a 2-torus and Partial Supersymmetry Restoration}
We saw in the previous subsection that the deformed geometry with a noncompact rigid 2-cycle generically breaks SUSY when $j_1 \ne j_2$ in (\ref{cccc1}). We now argue that the procedure employed in the resolved conifold geometry (extra compactification on  
$T^2$ with a magnetic flux on the torus) to preserve SUSY can also be applied after the geometric transition.  The geometry becomes a deformed conifold with an extra non-compact 2-cycle, fibered over a two torus. The deformation cycle $S^3$, its Poincare dual $P$  and the noncompact two cycle are all uplifted to four cycles.  

In the fourfold language, the identification (\ref{gluino}) becomes
\be
S = \int_{D_1} \Omega_{4}
\ee
where $\Omega_{4}$ is the holomorphic 4-form on the Calabi-Yau 4-fold and $D_1$ is a four-cycle  
$S^3 \times S^1_i$ where $S^1_i, i=1,2$ are the 1-cycles of the two torus.  The flavour degrees of freedom live on a noncompact 4-cycle $B_i$ which is an holomorphic noncompact 2-cycle fibered over $T^2$.  

To get the flux contribution to the effective superpotential,  we use the results of \cite{keshavsethi} for the  superpotential in case of a compactification on a Calabi-Yau 4-fold with a non-zero four form flux:
\begin{equation}
\label{4fold}
W = \int_{Y} \Omega \wedge G
\end{equation}
where $\Omega$ is a holomorphic four form on $Y$ and $G$ is an integral four form. This can be reduced to 
Calabi-Yau three folds by considering the Calabi-Yau fourfold as an elliptic fibre over a base $B$ and expanding
 $G$ as in \cite{keshavsethi}
\begin{equation}
\label{reduction}
G = q + p \wedge \chi + \sum_{i} H_i \wedge \theta^i
\end{equation}
where $q, p $ and $H_i$ are forms of degree 4, 2 and 3 on the base $B$, $\theta^i, i=1,2$ form a basis of integral one-forms on the fiber and $\chi$ is an integral two-form generating the two-dimensional cohomology of the elliptic fiber.  

In our case, the  Calabi-Yau fourfold is a deformed conifold fibered over a two torus. 
How do we incorporate the FI terms into the effective superpotential on the fourfolds? 
 When compactifying on a Calabi-Yau 3-fold, \cite{aga1} has argued that the  FI terms transform as 
a vector $(E_1, E_2, E_3)$ under an $SU(2)_R$ R-symmetry and, when considered as entries of a $2 \times 2$ matrix, the action appears as
\begin{equation}
\frac{1}{4 \pi} \mbox{Re(Tr} X \bar{E}). 
\end{equation}
$X$ is a $2 \times 2$ matrix depending on $j$ and $b_{NS}$ and $E_i$ are integrals over the P cycle:
\begin{equation}
\label{defe}
E_1 = \int_{P} \frac{H_{NS}}{g_s},~~E_2 = \int_P H_{RR},~~E_3 = \int_{P} d J/g_s
\end{equation}
We know from \cite{aga1} that the relevant supersymmetry variations of the $SU(2)_R$ doublets of fermions 
$(\psi,\lambda)$ are given by 
\be
\delta \Psi^i = X^{ij} \epsilon_j~~i,j=1,2
\ee
In case of several gauge groups with different values for $j$, there are several matrices $X_a$ with zero eigenvalues but the supersymmetry is broken for a  configuration with arbitrary values for $j_a$. 
 
To partially preserve supersymmetry, we consider the deformed conifold fibered over the two torus.  
We want to uplift the quantities $E_i$ to the Calabi-Yau fourfold. The 
term $H_i \wedge \theta^i$ in (\ref{reduction}) contains $\theta^i$, the basis for one-forms on the torus
\begin{equation}
d z = d x_1 + \tau d x_2,~~d \bar{z} = d x_1 + \bar{\tau} d x_2
\end{equation}
We consider a vector potential 
$A_2 = M x^1$. The noncompact 3-cycle $P$ becomes a collection of two $S^1_i \times P, i = 1,2$ four cycles denoted as  $\tilde{D}_1, \tilde{D}_2$. 
The uplift to the fourfold is
\be
\label{gauge}
\mbox{gauge coupling constant}:~\int_{B} H_{NS}   \leftrightarrow  \int_{\tilde{D}_1} H_{NS} \wedge  d x^1, 
\ee
\be
E_1 = \int_{B} \frac{H_{NS}}{g_s}  \leftrightarrow \tilde{E}_1 = \int_{\tilde{D}_1} H_{NS} \wedge d x^1,
\ee
\be
E_2 = \int_{B} H_{RR}  \leftrightarrow \tilde{E}_2 = \int_{\tilde{D}_1} H_{RR} \wedge d x^1,
\ee
\be
E_3 = \int_{B} d J/g_s \leftrightarrow \tilde{E}_3 = \int_{\tilde{D}_1} d J/g_s \wedge d x^1. 
\ee
On the other hand, we also have contributions from $\tilde{D}_2$ as 
\be
\int_{\tilde{D}_2} H_{NS} \wedge  A_2 d x^2, \int_{\tilde{D}_2} d J \wedge  A_2 d x^2, \int_{\tilde{D}_2} H_{RR} \wedge  A_2 d x^2
\ee

More involved is the calculation of the matrices $X_a$ for the deformed conifold fibration over the two torus. To do this, we need to split the 4-dimensional fermions into right (left) moving fermions on $R^{1,1}$. The definition of $X$ would contain both $j$ and the magnetic flux $M$. As mentioned before, in this work we restrict to the case when only flavours branes are charged under the magnetic flux and there is no $d J \ne 0$ on the 
$\tilde{D}_i$ cycles. All the information about the magnetic flux and nonzero D-terms is encoded in the noncompact 2-cycle. We plan to develop a general discussion for arbitrary $X$ and $E$ in a future publication.

We now return to our concrete example in this work and deal with the mismatch between $j_1 = 0$ and $j_2 \ne 0$ in 
(\ref{cccc1}). When lifted to the $SU(4)$ structure manifold, the phase introduced by the integral of $J$ over 
$\tilde{B}_2$ is matched by the phase introduced by $M$ on $T^2$. When integrated over the $B_2$, the $\tilde{B}_2$
fiber over $T^2$, the two phases cancel each other if the relation (\ref{relkuta}) is valid. 
For flavour branes, the geometric transition provides a set of cycles $D_i, \tilde{D}_i, B_2$.
The integral of $d J$ over $ \tilde{D}_i$ is zero when no D-term is considered for the gauge group. 
For magnetic flux on the two torus that only the flavours are charged under, the SUSY condition for D5 branes wrapped on a noncompact 4-cycle in the deformed conifold side is identical to the one in the resolved conifold side and requires the condition (\ref{relkuta}) to be true. We conclude that 
the condition (\ref{relkuta}) ensures that the SUSY is preserved during the geometric transition. 

As the geometries discussed here have $SU(4)$ structure in the presence of NS flux on Calabi-Yau fourfolds, it would be interesting to consider the approach of \cite{lustts} involving manifolds with 
SU(4) structure. This would allow us to consider more involved assignments of flavour and colour charges under the magnetic flux.

\section{Conclusions}

In this work, our goal was to provide a geometric picture for a partial supersymmetry breaking yielding (0,2) two dimensional theories. Our setup is T-dual to the brane configurations of \cite{kuta1}. We start with D5 branes wrapped on various compact or noncompact 2-cycles of resolved conifold geometries which are further fibered over a two torus.  Consequently, the D5 branes wrap four cycles $S$ which are (compact or non-compact) 2-cycles fibered over the two torus. The supersymmetry is partially preserved for rigid  two cycles and when a magnetic flux is considered on the two torus, if the magnetic flux and the rigidity parameter $j$ are related as in  
(\ref{relkuta}). This reproduces the equality between the D-terms and the magnetic fluxes proposed in 
\cite{kuta1}. We also consider the supersymmetry preservation after a geometric transition. For the case discussed in this paper, the supersymmetry condition involves  noncompact 2-cycles wrapped by flavour branes which are fibered over the two torus with magnetic flux. The relative phase between the central charges of various stacks of branes is zero when the relation (\ref{relkuta}) is obeyed.  

\vskip .5cm

{\bf{Note}}: As we were preparing to submit our results, a paper \cite{franco} appeared which considers  2d (0,2) quiver gauge theories on the worldvolume of D1-branes probing singular toric Calabi-Yau 4-folds. Our approach uses D5 branes wrapped on 4-cycles inside Calabi-Yau 4-folds and $SU(4)$ structure manifolds.

\vskip 1cm

\section*{Acknowledgments}

This work was supported in part by STFC under contract ST/L000431/1. We would like to thank Keshav Dasgupta for important comments on the draft.


\newpage

\begin{thebibliography}{99}

\bibitem{vafa}
C.~Vafa,
  ``Superstrings and topological strings at large N,''
  J.\ Math.\ Phys.\  {\bf 42}, 2798 (2001)
  [hep-th/0008142].

\bibitem{vafa1}
  F.~Cachazo, K.~A.~Intriligator and C.~Vafa,
  ``A Large N duality via a geometric transition,''
  Nucl.\ Phys.\ B {\bf 603}, 3 (2001)
  [hep-th/0103067].
F.~Cachazo, S.~Katz and C.~Vafa,
  ``Geometric transitions and N=1 quiver theories,''
  hep-th/0108120.
 F.~Cachazo, B.~Fiol, K.~A.~Intriligator, S.~Katz and C.~Vafa,
  ``A Geometric unification of dualities,''
  Nucl.\ Phys.\ B {\bf 628}, 3 (2002)
  [hep-th/0110028].


\bibitem{dva} R.~Dijkgraaf and C.~Vafa,
  ``On geometry and matrix models,''
  Nucl.\ Phys.\ B {\bf 644}, 21 (2002)
  [hep-th/0207106].
R.~Dijkgraaf and C.~Vafa,
  ``On geometry and matrix models,''
  Nucl.\ Phys.\ B {\bf 644}, 21 (2002)
  [hep-th/0207106].
 R.~Dijkgraaf and C.~Vafa,
  ``A Perturbative window into nonperturbative physics,''
  hep-th/0208048.

\bibitem{aga2}
M.~Aganagic, C.~Beem, J.~Seo and C.~Vafa,
  ``Geometrically Induced Metastability and Holography,''
  Nucl.\ Phys.\ B {\bf 789}, 382 (2008)
  [hep-th/0610249].

\bibitem{aga1} 
  M.~Aganagic and C.~Beem,
  ``Geometric transitions and D-term SUSY breaking,''
  Nucl.\ Phys.\ B {\bf 796}, 44 (2008)
  [arXiv:0711.0385 [hep-th]].

\bibitem{rt} 
  K.~Dasgupta, K.~Oh and R.~Tatar,
  ``Geometric transition, large N dualities and MQCD dynamics,''
  Nucl.\ Phys.\ B {\bf 610}, 331 (2001)
  [arXiv:hep-th/0105066],
  ``Open/closed string dualities and Seiberg duality from geometric
  transitions in M-theory,''
  JHEP {\bf 0208}, 026 (2002)
  [arXiv:hep-th/0106040].
  K.~h.~Oh and R.~Tatar,
  ``Duality and confinement in N = 1 supersymmetric theories from geometric
  transitions,''
  Adv.\ Theor.\ Math.\ Phys.\  {\bf 6}, 141 (2003)
  [arXiv:hep-th/0112040].
  R.~Roiban, R.~Tatar and J.~Walcher,
  ``Massless flavor in geometry and matrix models,''
  Nucl.\ Phys.\ B {\bf 665}, 211 (2003)
  [arXiv:hep-th/0301217];
K.~Landsteiner, C.~I.~Lazaroiu and R.~Tatar,
  ``Chiral field theories from conifolds,''
  JHEP {\bf 0311}, 057 (2003)
  [arXiv:hep-th/0310052]~~
K.~Landsteiner, C.~I.~Lazaroiu and R.~Tatar,
  ``(Anti)symmetric matter and superpotentials from IIB orientifolds,''
  JHEP {\bf 0311}, 044 (2003)
  [arXiv:hep-th/0306236]~~
 R.~Tatar and B.~Wetenhall,
  ``Metastable vacua, geometrical engineering and MQCD transitions,''
  JHEP {\bf 0702}, 020 (2007)
  [arXiv:hep-th/0611303].

\bibitem{gu} 
H.~Garcia-Compean and A.~M.~Uranga,
  ``Brane box realization of chiral gauge theories in two-dimensions,''
  Nucl.\ Phys.\ B {\bf 539}, 329 (1999)
  [hep-th/9806177].

\bibitem{guko1}
A.~Gadde, S.~Gukov and P.~Putrov,
  ``Fivebranes and 4-manifolds,''
  arXiv:1306.4320 [hep-th].

\bibitem{guko2}
 A.~Gadde, S.~Gukov and P.~Putrov,
  ``(0, 2) trialities,''
  JHEP {\bf 1403}, 076 (2014)
  [arXiv:1310.0818 [hep-th]].
A.~Gadde, S.~Gukov and P.~Putrov,
  ``Exact Solutions of 2d Supersymmetric Gauge Theories,''
  arXiv:1404.5314 [hep-th].
J.~Guo, B.~Jia and E.~Sharpe,
  ``Chiral operators in two-dimensional (0,2) theories and a test of triality,''
  arXiv:1501.00987 [hep-th].
  E.~Sharpe,
  ``A few recent developments in 2d (2,2) and (0,2) theories,''
  arXiv:1501.01628 [hep-th].

\bibitem{guko3}
N.~Bobev, K.~Pilch and O.~Vasilakis,
  ``(0, 2) SCFTs from the Leigh-Strassler fixed point,''
  JHEP {\bf 1406}, 094 (2014)
  [arXiv:1403.7131 [hep-th]].

\bibitem{guko4}
B.~Jia, E.~Sharpe and R.~Wu,
  ``Notes on nonabelian (0,2) theories and dualities,''
  arXiv:1401.1511 [hep-th].

\bibitem{kuta1} 
  D.~Kutasov and J.~Lin,
  ``(0,2) Dynamics From Four Dimensions,''
  Phys.\ Rev.\ D {\bf 89}, 085025 (2014)
  [arXiv:1310.6032 [hep-th]].

\bibitem{kuta2} 
  D.~Kutasov and J.~Lin,
  ``(0,2) ADE Models From Four Dimensions,''
  arXiv:1401.5558 [hep-th].

\bibitem{franco} 
  S.~Franco, D.~Ghim, S.~Lee, R.~K.~Seong and D.~Yokoyama,
  ``2d (0,2) Quiver Gauge Theories and D-Branes,''
  arXiv:1506.03818 [hep-th].


\bibitem{hw}
A.~Hanany and E.~Witten,
  ``Type IIB superstrings, BPS monopoles, and three-dimensional gauge dynamics,''
  Nucl.\ Phys.\ B {\bf 492}, 152 (1997)
  [hep-th/9611230].

\bibitem{uc} 
  J.~F.~G.~Cascales and A.~M.~Uranga,
  ``Branes on generalized calibrated submanifolds,''
  JHEP {\bf 0411}, 083 (2004)
  [hep-th/0407132].


\bibitem{moore}
M.~Marino, R.~Minasian, G.~W.~Moore and A.~Strominger,
  ``Nonlinear instantons from supersymmetric p-branes,''
  JHEP {\bf 0001}, 005 (2000)
  [hep-th/9911206].

\bibitem{mn}
J.~M.~Maldacena and C.~Nunez,
  ``Towards the large N limit of pure N=1 superYang-Mills,''
  Phys.\ Rev.\ Lett.\  {\bf 86}, 588 (2001)
  [hep-th/0008001].


\bibitem{ks}
I.~R.~Klebanov and M.~J.~Strassler,
  ``Supergravity and a confining gauge theory: Duality cascades and chi SB resolution of naked singularities,''
  JHEP {\bf 0008}, 052 (2000)
  [hep-th/0007191].

\bibitem{minagra}
  A.~Butti, M.~Grana, R.~Minasian, M.~Petrini and A.~Zaffaroni,
  ``The Baryonic branch of Klebanov-Strassler solution: A supersymmetric family of SU(3) structure backgrounds,''
  JHEP {\bf 0503}, 069 (2005)
  [hep-th/0412187].

\bibitem{ds1} 
  A.~Dymarsky, I.~R.~Klebanov and N.~Seiberg,
  ``On the moduli space of the cascading SU(M+p) x SU(p) gauge theory,''
  JHEP {\bf 0601}, 155 (2006)
  [hep-th/0511254].

\bibitem{mm} 
  J.~Maldacena and D.~Martelli,
  ``The Unwarped, resolved, deformed conifold: Fivebranes and the baryonic branch of the Klebanov-Strassler theory,''
  JHEP {\bf 1001}, 104 (2010)
  [arXiv:0906.0591 [hep-th]].

\bibitem{cdt} 
  F.~Chen, K.~Dasgupta, P.~Franche, S.~Katz and R.~Tatar,
  ``Supersymmetric Configurations, Geometric Transitions and New Non-Kahler Manifolds,''
  Nucl.\ Phys.\ B {\bf 852}, 553 (2011)
  [arXiv:1007.5316 [hep-th]].


\bibitem{kem} 
  K.~Dasgupta, M.~Emelin and E.~McDonough,
  JHEP {\bf 1502}, 179 (2015)
  [arXiv:1412.3123 [hep-th]].

\bibitem{int1} 
  K.~Intriligator, H.~Jockers, P.~Mayr, D.~R.~Morrison and M.~R.~Plesser,
  ``Conifold Transitions in M-theory on Calabi-Yau Fourfolds with Background Fluxes,''
  Adv.\ Theor.\ Math.\ Phys.\  {\bf 17}, 601 (2013)
  [arXiv:1203.6662 [hep-th]].


\bibitem{becker}
  K.~Becker, M.~Becker and A.~Strominger,
  ``Five-branes, membranes and nonperturbative string theory,''
  Nucl.\ Phys.\ B {\bf 456}, 130 (1995)
  [hep-th/9507158].

\bibitem{becker1}
K.~Becker, M.~Becker, D.~R.~Morrison, H.~Ooguri, Y.~Oz and Z.~Yin,
  ``Supersymmetric cycles in exceptional holonomy manifolds and Calabi-Yau 4 folds,''
  Nucl.\ Phys.\ B {\bf 480}, 225 (1996)
  [hep-th/9608116].


\bibitem{lust} 
  D.~Lust, P.~Mayr, R.~Richter and S.~Stieberger,
  Nucl.\ Phys.\ B {\bf 696}, 205 (2004)
  [hep-th/0404134].
 
\bibitem{mcg}
A.~Lawrence and J.~McGreevy,
  ``Local string models of soft supersymmetry breaking,''
  JHEP {\bf 0406}, 007 (2004)
  [hep-th/0401034].
M.~Grana, J.~Louis and D.~Waldram,
  ``Hitchin functionals in N=2 supergravity,''
  JHEP {\bf 0601}, 008 (2006)
  [hep-th/0505264].

\bibitem{agava2}
 M.~Aganagic and C.~Vafa,
  ``Mirror symmetry, D-branes and counting holomorphic discs,''
  hep-th/0012041.

\bibitem{keshavsethi}
 S.~Gukov, C.~Vafa and E.~Witten,
  ``CFT's from Calabi-Yau four folds,''
  Nucl.\ Phys.\ B {\bf 584}, 69 (2000)
  [Erratum-ibid.\ B {\bf 608}, 477 (2001)]
  [hep-th/9906070];
K.~Dasgupta, G.~Rajesh and S.~Sethi,
  ``M theory, orientifolds and G - flux,''
  JHEP {\bf 9908}, 023 (1999)
  [hep-th/9908088].

\bibitem{lustts} 
D.~Lust, P.~Patalong and D.~Tsimpis,
  ``Generalized geometry, calibrations and supersymmetry in diverse dimensions,''
  JHEP {\bf 1101}, 063 (2011)
  [arXiv:1010.5789 [hep-th]].
D.~Prins and D.~Tsimpis,
  ``IIB supergravity on manifolds with SU(4) structure and generalized geometry,''
  JHEP {\bf 1307}, 180 (2013)
  [arXiv:1306.2543 [hep-th]].

\end{thebibliography}
\end{document}